\documentstyle[psfig]{./elsart}

\begin{document}
\begin{frontmatter}
\title{The effect of continuum scattering processes on
spectral line formation}
\author{G. Israelian}
\address{Insituto de Astrof\'\i sica de Canarias, E-38200 La Laguna, 
Tenerife, Spain}

\begin{abstract}

The effect of scattering processes in the continuum
on the formation of spectral lines in a static  
atmosphere with an arbitrary distribution of the internal energy
sources is investigated using Ambartsumian's principle of
invariance. Spectral line profiles are calculated to
illustrate the effect the assumption of the complete
redistribution on atoms and coherent scattering in
continuum may have on the emergent intensity. 
The one-dimensional case is considered for simplicity.

\end{abstract}

\end{frontmatter}

\section{Introduction}

In the study of the formation of spectral lines, an interesting
problem is the transfer of radiation in a medium consisting
of centers at which scattering occurs for all frequencies.
This class of problems includes in particular that of the
formation of spectral lines in one-dimensional, semi-infinite
media.   
In his pionering paper Schuster$^{1}$ showed that under some 
circumstances scattering of photons in continuum can lead
to the formation of emission lines in the spectrum. The formation 
mechanism of emission lines due to scattering in continuum is
simple. A strong scattering source in the continuum may force
the photons to be more scattered than absorbed in the certain part
of the spectrum and will therefore  lead to a decrease in the source function
in the continuum. Suppose that the radiation field  
in a spectral line is in local thermodynamical equilibrium (LTE) 
with the local medium so that the line photons have much more chance
of being absorbed and destroyed in the medium than  photons 
in the continuum. The increase in the efficiency of scattering causes  
the radiation field in the continuum to decrease in the meantime, thereby
keeping the flux 
in the spectral line unaffected  because of the thermal 
radiation of the medium. Thus the spectral line will appear in 
emission. 

\section{Spectral lines formed in an isothermal medium}

Recently Israelian and Nikoghossian$^{2}$ investigated 
the diffusion of isotropic and coherent radiation in a semi-infinite 
medium by the probabilistic approach. The probabilistic
method is based on the concept of a probability of quantum exit from 
a medium  and has been successfully
applied to treat many classical problems in a theory of
radiative transfer$^3$.
The principle of optical reversibility, together with 
the principle of invariance, allows  very complex
problems to be handled when the probabilistic approach is employed. 
  
Let us consider a one-dimensional, semi-infinite, isothermal
medium with a distribution of energy sources:
 
\begin{equation}
\epsilon (\tau , x) = u(x)B[T(\tau)],
\end{equation}

where $u(x) = (1 - \lambda)\alpha(x) + \beta $. Here we denote
$x$ as a non-dimensional frequency 
($x = \delta \nu / \delta \nu_{D}$, where 
$\delta \nu_{D}$ is a Doppler width), $\alpha(x)$ as the normalized 
absorption coefficient profile in the line, $B[T(\tau)]$ 
is the Planck function, which depends on the optical depth, $\tau$,
relative to the central wavelength of the line through the 
temperature, $T$, $\lambda$ is the probability of 
photon re-emission, $\beta$ is the ratio of the absorption
coefficient in the continuum to that at the center of the 
spectral line. We have also introduced$^2$
a parameter $\gamma$, the ratio of the scattering coefficient
in the continuum to the absorption coefficient at the line center.
Let us assume that the medium has an infinitely large optical
thickness so that the addition of a layer of small optical thickness
will not change the reflection coefficient of the medium.
This {\it principle of invariance} of Ambartsumian$^{4,5}$
 allows one to
obtain the  following equation:

\begin{eqnarray}
\nonumber
\frac{2}{\lambda}[v(x') + v(x)] \rho(x,x') = r(x,x') 
+ \int r(x',x'')\rho(x'',x)dx'' \\
+ \int r(x'',x)\rho(x',x'')dx'' + 
  \int \rho(x',x'')dx''\int r(x'',x''')\rho(x''',x)dx''',
\end{eqnarray}

where $v(x) = \alpha(x) + \beta + \gamma$, $\rho(x,x')$ and 
$r(x,x')$ 
are the reflection and redistribution functions, respectively.
The reflection function, $\rho(x,x')$, has the following 
probabilistic meaning: if a photon of frequency $x'$ is 
incident on the medium, then $\rho(x,x')dx$ is the probability
that, after multiple scattering, a photon is reflected 
from the medium with a frequency in the interval $x$,$x+dx$.

Hereafter, all integrations over $x$ will be carried out from 
$- \infty$ to $+ \infty$. In general, there can be several 
sources of absorption/scattering in the medium, in which
case $r(x,x')$ must be considered as a sum of redistribution
functions due to different centers (like atoms, electrons, dust
particles, molecules, plasmons, etc.). 
Let us also add pure scattering centers
in the medium and denote their redistribution function as
$r_{e}(x,x')$. If these  centers are free electrons and 
the scattering is incoherent$^5$ 
then

\begin{equation}
r_{e}(x,x') ={\delta \nu_{D}}^{-1}\pi^{-0.5}(e^{-x^{2}} - 
2x\pi\int_{x}^{\infty} exp(-u^{2})du).
\end{equation} 

In case of the coherent scattering we have:

\begin{equation}
r_{e}(x,x') = \frac{\gamma}{\lambda} \delta(x - y).
\end{equation}

The total redistribution function in the case of the
complete redistribution of photons on the atoms and
the coherent scattering on the electrons will be 

\begin{equation}
r(x,x') = \alpha(x)\alpha(x') + 
\frac{\gamma}{\lambda} \delta(x - y).
\end{equation}

The problem of determining the radiation field in a 
medium with or without energy sources under broad 
assumptions concerning the elementary scattering 
process can be simplified if the redistribution
function for atoms allows  presentation in the form of a 
bilinear expansion$^{6,7,8}$:

\begin{equation}
r(x,x') = \sum_{k=0}^{\infty} A_{k}\alpha_{k}(x)\alpha_{k}(x'),
\end{equation} 
where 
\begin{eqnarray*}
A_{k}=\frac{1}{2k+1}, \; \alpha_{k}(x)=
\frac{1}{\pi^{0.25}2^{k}\sqrt{(2k)}!}e^{-x^{2}}H_{2k}(x)
\end{eqnarray*}

and $H_{2k}(x)$ are the Hermit polynomials.
Note that with only the first term taken in the expansion (6),
we arrive at complete frequency redistribution, i.e., there
is no correlation between the frequency of the incoming 
and  the scattered quantum in the observer's frame of
refernce.
Such a representation can be obtained by expanding the
function $r(x,x')$ with respect to its eigenfunctions over
($-\infty$,$\infty$). In general, the function $r(x,x')$ 
can be replaced 
by an finite sum of its expansion with respect to its 
eigenfunctions over ($-\infty$,$\infty$):

\begin{equation}
r(x,x') \approx r_{N}(x,x')=\sum_{k=0}^{N} \frac{\chi_{k}(x)\chi_{k}(x')}
{\lambda_{k}},
\end{equation}
where the functions $\chi_{k}(x)$ are the normalized solutions of
the equations

\begin{equation}
\chi_{k}(x) = \lambda_{k}\int_{-\infty}^{\infty}
r(x,x')\chi_{k}(x')dx'.
\end{equation}

For fixed $N$, this partial sum gives the best mean-square solution
approximation in the ($x',x$)-plane to the function
$r(x',x)$ among all possible representations. In the case
of representation (7), the accuracy of the fulfillment
of  condition (8) agrees with the accuracy of the
approximation of the function $r(x,x')$ by the partial
sum $r_{N}(x,x')$. Different methods 
(like the method of least squares, the method of moments, etc.)
exist for the approximate
construction of the eigenfunctions of the kernel $r(x,x')$ and
we shall not dwell on this point.

Suppose that the function $r_{e}(x,x')$ can be expanded as well:

\begin{equation}
r_{e}(x,x') = \sum_{m=0}^{\infty} B_{m}\omega_{m}(x)\omega_{m}(x'). 
\end{equation}

Then in a more general case we will have a sum of functions
(6) and (9). 
If the sum of representations (6) and (7) for the redistribution
function, $r(x,x')$, is used, then the integral equation (2)
can be rewritten as:

\begin{equation}
\rho(x,x')=\frac{\lambda}{2}\left[\sum_{k=0}^{\infty}A_{k}\frac{\varphi_{k}(x)\varphi_{k}(x')}
{v(x) + v(x')} +
\sum_{i=0}^{\infty}B_{i}{\frac{\psi_{i}(x)\psi_{i}(x')}
{v(x) + v(x')}}\right],
\end{equation}

where the functions $\varphi_{k}$ and $\psi_{i}(x)$ 
satisfy the set of functional equations providing a
generalization of Ambartsumian's functional 
equation for non-coherent scattering:

\begin{eqnarray}
\varphi_{k}(x) &=& \alpha_{k}(x) + \int_{-\infty}^{\infty}
\alpha_{k}(z)\rho(x,z)dz,\\
\psi_{i}(x)    &=& \omega_{i}(x) + \int_{-\infty}^{\infty}
\omega_{i}(x)\rho(x,z)dz.
\end{eqnarray}

The solution of this system provides a reflection
function, $\rho(x,x')$, which allows the evaluation of the
profiles of spectral lines (since $R(x)$ = $\int\rho(x,x')dx'$). 
Determination of the profiles
of absorption lines in an isothermal atmosphere
is equivalent to that for diffuse reflection of radiation
from a semi-infinite medium$^3$.
The reflection and absorption profiles in this case
are related by

\begin{equation}
R(x) + \tilde{R}(x) = 1.
\end{equation}

This relation enables one to reduce line-profile calculations
for an isothermal atmosphere to the evaluation of the 
reflection function, $\rho(x,x')$.  

In this article we present numerical results for 
the  simplest case of complete
redistribution of photons on atoms and a 
coherent scattering on free electrons. 
If last term in the equation (5) describes a coherent
scattering in the continuum, then it can be written
in the form

\begin{equation}
\frac{\gamma}{\lambda}\delta(x - x') = 
\frac{\gamma}{\lambda}\sum_{k=0}^{\infty} \alpha_{k}(x')\alpha_{k}(x),
\end{equation}

using the set of the $\alpha_{k}(x)$ functions given
in the expansion (6).

\section{Spectral line formation in a non-isothermal medium}

In the case of a non-isothermal medium we shall assume that 
the source function can be presented in the following form:

\begin{equation}
B[T(\tau)] = \sum_{n=0}^{\infty} \frac{B_{n}}{n!} (\beta \tau)^{n},
\end{equation}
where $B_{n}$ is the $n^{\rm th}$ derivative of the Planck function.
Let us denote by $Y(\tau,x',x)$ the probability that a photon with a
frequency $x'$ moving in the medium at an optical depth $\tau$  
will leave it with a frequency in the range ($x, x + dx$). 
Similarly, let $P(\tau,x',x)$ be the same probability but for 
photons absorbed at an optical depth $\tau$. Application of 
the invariance principle and  simple physical reasoning results 
in the following equations:

\begin{eqnarray}
\frac{\partial Y(\tau,x',x)}{\partial \tau} &=& - v(x)Y(\tau,x',x) 
+ \int_{-\infty}^{\infty} P(0,z,x)\,\alpha(z)\,Y(\tau,x',z)\,dz \\
\alpha(x) P(0,x',x) &=& \frac{\lambda}{2} r(x',x) + 
\frac{\lambda}{2} \int_{-\infty}^{\infty} r(x',z)\,\rho(z,x)\,dz
\end{eqnarray}

These equations can be easily derived when $\beta$=0
(Sobolev$^{4}$) and $\beta \neq$0 
(Nikoghossian and Haruthyunian$^{8}$).

The intensity of the emergent radiation, $I(0,x)$, with the 
source function given by (15) can be presented in the
form

\begin{equation}
I(0,x)=\sum_{n=0}^{\infty}B_{n}I_{n}(0,x),
\end{equation}

where

\begin{equation}
I_{n}(0,x)=\int_{-\infty}^{\infty}u(x')dx'\int_{0}^{\infty}
f_{n}(\tau)Y_(\tau,x',x)d\tau
\end{equation}

and $f_{n}(\tau) = (\beta \tau)^{n}/n!$. 

The spectral line contours for the mentioned distribution
of the energy sources can be calculated with the formulae$^7$

\begin{equation}
v(x)R_{n}(x) = \int_{-\infty}^{\infty}
\alpha(x')P(0,x',x)R_{n}(x')dx'+ \beta R_{n-1}(x).
\end{equation}

Note that $\gamma$ appears only on the left-hand side of this
equation. The final profiles are

\begin{equation}
R(x) = \frac{\sum_{n}^{} B_{n} R_{n}(x)}{\sum_{n}^{} B_{n}}.
\end{equation}

Note that the function $\rho(x,x')$ in (17)
is the reflection function from a semi-infinite 
isothermic medium. Since we have presented this later in the
form of eq. (9), we can write, instead of eq. (17),

\begin{equation}
\frac{2}{\lambda}\alpha(x') P(0,x',x)= 
\sum_{k=0}^{\infty} A_{k}\alpha_{k}(x')\varphi_{k}(x) + 
\sum_{m=0}^{\infty} B_{m}\omega_{m}(x')\psi_{m}(x).
\end{equation}

Let us multiply eq. (20) by $\alpha_{k}(x)$ 
and integrate with respect to $x$ over the range ($-\infty$
$\infty$). Repeating the same operation for 
$\omega_{m}(x)$ yields, for 
$R_{n}^{k}=\int\alpha(x')R_{n}(x')dx$, the following system of 
the algebraic equations:

\begin{eqnarray}
R_{n}^{k} &=& \sum_{k=0}^{\infty}A_{k}R_{n}^{k}\xi_{k} + 
\sum_{m=0}^{\infty}B_{m}R_{n}^{m}\sigma_{k,m} + 
\delta_{k}^{(n-1)} \\
R_{n}^{m} &=& \sum_{k=0}^{\infty}A_{k}R_{n}^{k}\mu_{m,k} + 
\sum_{m=0}^{\infty}B_{m}R_{n}^{m}\theta_{m} + \zeta_{m}^{(n-1)},
\end{eqnarray}

where we have introduced the notation

\begin{eqnarray*}
\xi_{k} &=& \int_{-\infty}^{\infty}
\frac{\alpha_{k}(x)\varphi_{k}(x)}{v(x)}dx, \; 
\sigma_{k,m} = \int_{-\infty}^{\infty}
\frac{\alpha_{k}(x)\psi_{m}(x)}{v(x)}dx, \\
\delta_{k}^{(n)} &=& \beta\int_{-\infty}^{\infty}
\frac{\alpha_{k}(x)}{v(x)}R_{n}(x)dx, \;
\mu_{m,k} = \int_{-\infty}^{\infty}\frac{\omega_{m}(x)\varphi_{k}(x)}
{v(x)}dx, \\
\theta_{m} &=& \int_{-\infty}^{\infty}
\frac{\omega_{m}(x)\psi_{m}(x)}{v(x)}dx, \;
\zeta_{m}^{(n)} = \beta\int_{-\infty}^{\infty}
\frac{\omega_{m}(x)}{v(x)}R_{n}(x)dx.
\end{eqnarray*}

Apparently the question is reduced to solving the system 
of algebraic equations (23--24). It is obvious  
that we can add different absorption/scattering centers
(atoms, molecules, etc.) to the medium, and if their 
redistribution functions
allow bilinear expansions, the problem can be reduced to
a system of algebraic equations. For a medium containing two
kinds of absorbing/scattering centers, one needs to construct
and tabulate functions $\varphi_{k}(x)$ and $\psi_{m}(x)$. 
Once this has been done, one needs to compute the constants 
$\xi_{k},\sigma_{k,m},\delta_{k}^{(n)},$
$\mu_{m,k},\theta_{m}$, and $\zeta_{m}^{(n)}$, after which the
line profiles are evaluated from eqs. (20) and (21) 
using recursion relations.
This discussion  clearly remains valid for more than
two kinds of absorbing/scattering centers in the medium 
(provided that their redistribution functions allow a 
bilinear expansion) and for more general assumptions relative 
to the geometry of the medium.

\section{Discussion}

It is known that  coherent scattering in a medium gives
 the source function a non-LTE character. This fact
alone makes it possible for spectral lines to appear
in emission, even if the $line$ source function 
does not deviate from LTE$^1$.
The mechanism requires
a low temperature gradient, a strong spectral line and
an LTE process of line formation. The criteria derived
for the appearance of the lines in emission in a
non-isothermal stellar atmosphere are much more stringent and
it is possible that this mechanism cannot work in a
$real$ stellar atmosphere. Another difficulty in producing 
emission lines in a $normal$ atmosphere is the small 
cross-section of the Thomson scattering. 
However, in the general case, Thomson scattering on free
electrons is not necessarily the only nor the most
effective mechanism among the scattering processes
which may exist in astrophysical plasmas.  

We have computed line profiles from a semi-infinite,
one-dimensional atmosphere assuming complete 
redistribution on atoms and  coherent scattering
on free electrons. For the isothermal and non-isothermal
($n$=1) cases we have numerically solved the  systems of 
equations (10--12) and (23--24), respectively.
 Problems connected with the method of approximation
for the solution of the system of equations of the form
(10--12) are discussed in detail in the work of 
Haruthyunian and Nikoghossian$^{6}$.

Figures 1 and 2 illustrate 
the absorption/emission profiles for different values 
of $\gamma$, $\lambda$, and $\beta$. The profiles for the 
isothermal and non-isothermal ($n$=1) media are shown 
in Figs. 1 and 2, respectively. To compute the $B_{n}$ coefficients
in the sum (15) we have considered a hypothetical
line at 6000 \AA, a stellar atmosphere with $T_{\rm eff}$ =
40000 K and a gray-atmosphere temperature law.
We can see that even in the case of a non-LTE source function
in the line (i.e., $\lambda <$1) one can have emission lines
depending on the efficiency of scattering and absorption
in continuum. The five parameters involved in this 
problem are the scattering ($\gamma$) and the absorption
($\beta$) in continuum, the scattering ($\lambda$) and
the absorption ($\alpha(x)$) in the line and the 
temperature gradient in the medium. The addition of
scattering to the atmosphere can raise or lower the local
continuum depending on the location of the spectral
line, the efficiency of scattering and a temperature 
gradient$^{10}$. Therefore,
the variations of continuum in Figs. 1 and 2 have
 only a {\it local} character. We have already seen$^{10}$
that the variation of the continuum 
due to the scattering is nothing else than a 
$redistribution$ of emergent radiation demanded by the 
condition of radiative equilibrium 
and the conservation of integral flux. Our Fig. 2 clearly
demonstrates that the conditions for emission are much
more stringent in the non-isothermal than in the 
isothermal atmosphere. It is also possible to have a 
situation in which the wings (formed in LTE deep in the atmosphere)
of a strong line  will appear in emission while the line core 
is in absorption because of the strong non-LTE effects.
Interesting effects may appear when one studies a wing and
a core of the spectral line separately, considering depth 
and frequency dependence of $\gamma$, $\beta$, and $\lambda$.

\begin{ack}
The author is grateful to Prof. A. Nikoghossian and Dr. H.
Haruthyunian for the helpful discussions.
\end{ack}

\newpage

\begin{figure}
\centerline{\psfig{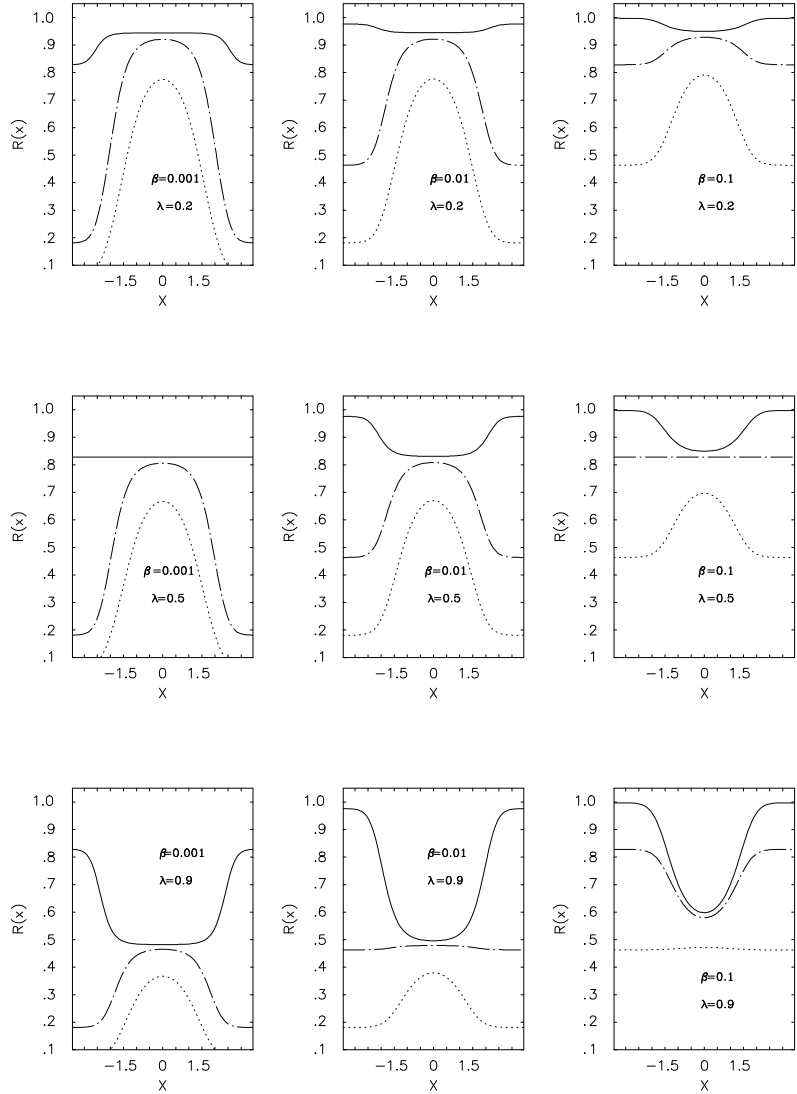}}
\caption[]{Spectral line profiles in the case of
complete frequency redistribution on atoms and
coherent scattering in continuum for an
isothermal atmosphere. Computations have been done
for $\gamma$=0.001 (solid line), $\gamma$=0.1 (dash-dotted line)
and $\gamma$=1.0 (dotted line). Values of $\beta$ and
$\lambda$ are indicated. 
}
\end{figure}

\begin{figure}
\centerline{\psfig{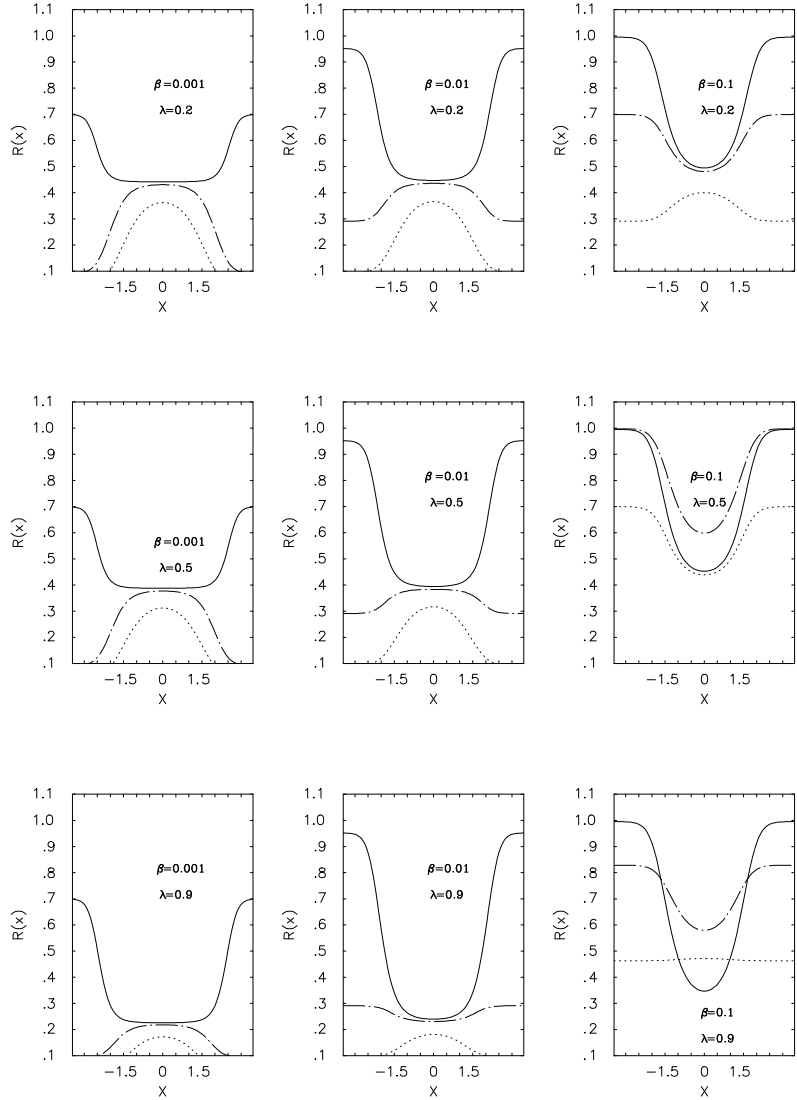}}
\caption[]{The same as Fig 1. but for a non-isothermal
atmosphere with $n=1$ (see text).}
\end{figure}

\end{document}